\documentclass[proceedings]{rmaa}

\title{Magellanic Clouds Planetary Nebulae: an updated view on stellar
evolution and populations}

\author{Letizia Stanghellini,\altaffilmark{1,2}
R. A. Shaw,\altaffilmark{1}, and  M. Mutchler,\altaffilmark{1}}
 
\altaffiltext{1}{Space Telescope Science Institute.}
\altaffiltext{2}{European Space Agency.}

\fulladdresses{
\item 3700 San Martin Drive
Baltimore, MD 21218 (USA).}

\shortauthor{STANGHELLINI ET AL.}
\shorttitle{Magellanic Cloud PNs}

\abstract{Planetary Nebulae (PNs) in the Magellanic Clouds are studied
to understand stellar populations and evolution of low- and 
intermediate-mass stars in different chemical environments.
Using {\it HST} observations from our LMC and SMC PN morphological survey and 
from the {\it HST} Data Archive, we look at the relations between PN morphology and
their evolution and populations. In this paper we show some of our 
recent results on these relations, in an historical context.}


\begin{document}

\maketitle

\section{Open questions on Planetary Nebula Morphology and Evolution}

During the past decades, significant progress has been made 
toward the understanding of the late phases of stellar 
evolution of low- and intermediate-mass stars through the
study of Galactic and Extra-galactic Planetary Nebulae (PNs).
Planetary Nebulae are the gaseous relics of the evolution of
stars in the 1--8 M$_{\odot}$ mass range, and they carry a wealth of 
information on the physical status of their progenitors. 

The morphology of PNs is an essential physical parameter to
know in order to construct reliable models of nebulae and
stars. Pioneering studies on the connections between PN morphology
and the evolution of the stars, and the progenitor populations,
started in the early 1970s with the work of
Greig (1972), and continued with 
Peimbert (1978), and Peimbert and Torres--Peimbert (1983).
The work done by the Peimbert's group set the stage for
more recent studies, by showing that PNs in the Galaxy are the progeny of
different stellar populations, and that there is a connection between the
scale height distribution of Galactic PNs and their chemical content. In fact,
Peimbert (1978) found that 
PNs in the Galactic disk are underabundant in carbon, and overabundant 
in nitrogen, as expected from stellar evolution of stars more massive than
about 2--3 M$_{\odot}$. 
Morphology is another link between Galactic spatial distribution and
chemistry of PNs: the nitrogen overabundant PNs in the Galactic disk
tend to be Bipolar in shape.

The early results have been later confirmed on the basis of larger
databases. Stanghellini et al. (1993) suggested that
the Bipolar nebulae host central stars with higher masses than those
hosted by Elliptical
and  Round nebulae. 
Manchado et al. (2000) performed a similar
analysis with data in the
IAC catalog of northern PNs (Manchado et al. 1996). The analysis basically
confirms the earlier results,
(Manchado et al. 2000)  with the added bonus of the reliability of a complete
and homogeneous PN sample.

Despite the large amount of important work on planetary nebulae and their
evolution, there are still fundamental questions that
remain unanswered. The main question related to PN morphology is
the quest for the mechanism(s) that produce symmetric or asymmetric PNs, and
how the appropriate mechanisms can account for the observed scenarios.
An excellent description of the present understanding of the link between 
morphology and evolution has been presented by Guillermo Garcia--Segura 
at this conference (Garcia--Segura et al., this volume). It is more and
more evident that measuring absolute physical parameters of
PNs is essential in order to constraint the models.

A thorough analysis and modeling of the Magellanic Cloud PNs is 
very important to achieve the next level of
understanding of PNs and their environment. Toward this end, it is essential to
minimize the distance
(and stellar luminosity) uncertainties. Studies of
PNs in the Magellanic Clouds are of fundamental importance in answering 
the next set of open questions:

\begin{enumerate}
\item{Is PN morphology related to the progenitor Populations, and
its chemistry?}
\item{How does PN morphology relate to the galaxian
properties, such as metallicity, star formation history, and  
distribution of stellar populations?}
\end{enumerate}

\section{The importance of Magellanic Cloud Planetary Nebulae}

\subsection{Background}

The spatial resolution achieved with {\it HST} observations 
allows to explore Extra-galactic PN morphology, and its 
relations to the physics of the central stars and to nebular evolution. 
Planetary Nebulae in the Magellanic Clouds are exceptionally suited for 
this type of studies, for their known distances and their low field
reddening.

Since the mid 1950s, planetary nebulae have been identified and
confirmed spectroscopically in the Magellanic Clouds (e. g., Lindsay 1955;
Webster 1969). The Magellanic Cloud (MC) PNs are typically spatially unresolved
from the ground. For this reason, the earliy science results are mainly
based on PN spectra, used to determine the
nebular abundances, the plasma diagnostics, 
and the motion of the PNs within the host galaxies 
(Dopita et al. 1985; Aller \& Keyes 1987; Peimbert 1987; Boroson \& Liebert 
1989; Barlow 1991; Torres-Peimbert 1993).

The inability to resolve the MC PNs spatially with 
ground-based astronomy did not 
prevent studying the correlations between the PNs and their central stars,
using stellar physical parameters inferred from the physics of the
host nebulae.
Kaler \& Jacoby (1990) found correlations between the central star masses and
the N/O and C/O abundances of the PN shells in the MC. Their technique was
to determine stellar temperature via the cross--over temperature method,
then get the central star masses from their location on the HR diagram.
Among other findings, this work revealed the nitrogen depletion 
of the low-mass central stars. 

The availability of the {\it HST} imagery naturally induced a renassance of PN 
studies in the Magellanic Clouds.
Even with the earlier, optically-aberrated {\it HST} observations, 
studies of LMC and SMC PNs led to morphological insight
(Blades et al. 1992; Dopita et al. 1996; 
Vassiliadis et al. 1998; Stanghellini et al. 1999). As WFPC2 and STIS 
became available on {\it HST}, the level of accuracy 
of the morphological studies of MC PNs reached the level of detail
of Galactic PN morphology via ground-based observations.

\subsection{Our HST/STIS Morphological Program}

In this paper, we show the early results of a large project 
on the morphology, evolution, and populations of the Magellanic Clouds
PNs. Our project aims at studying the correlations between morphology
and stellar evolution and populations in PNs in
the Magellanic Clouds.
To this end, we use our own {\it HST} observations,
as well as data from the {\it HST} Data Archive. The archived data consists of
MC PN images including 29 PNs. Most of these images 
are compromised by spherical aberration, thus their resolution is limited.
Our own {\it HST} images are acquired within two {\it HST} snapshot
programs: 8271 (LMC PNs) and 8663 (SMC PNs). The LMC program is
complete with 29 PN observed to date. The SMC program
is still active, counting so far about 20 PNs. 
Here we describe the results
obtained from the LMC program, but the importance of the SMC PN observations
and a preview of the early results is also discussed.

Our MC PNs are observed with STIS slitless spectroscopy (Shaw et al. 2001). 
This method produces a series of {\it narrowband}
images in the prominent nebular lines, and achieve spatial and 
spectral resolution at once. Together with the slitless dispersion, a 
{\it clear filter} image is also taken for each nebula, as a reference
for the general morphology and to locate the central star, if
visible, and to determine its magnitude. 
The archived PN images, together with  our newly observed images, where
classified morphologically as Round (R), Elliptical (E), Bipolar (B),
Bipolar Core (BC), Quadrupolar (Q), and Pointsymmetric (P). All types
are discussed  in Manchado et al. (1996), except the BC class, discussed
by Stanghellini et al. (1999). The spatial resolution that we achieved
with the STIS images of LMC PNs is as good as the typical ground based 
resolution for Galactic PNs that are 3 kpc away. This means that
we can acquire a reliable morphological classification of
our sample MC PNs.
Nonetheless, it is impossible with the existing technology
to detect the fine details in MC PNs. In particular,
the pointsymmetry, when associated to the Bipolarity, can be hard to
detect.

We found that the individual morphological types in LMC PNs are 
similar to those of Galactic PNs. 
In Figure 1 we show a sampler of LMC PNs from our survey.
We show the images as observed through the clear filter, and the [O III] 
$\lambda$5007 \AA~ contour plots. 
We found in our sample  Round, Elliptical (e.g., SMP 4), Bipolar
(e.g., SMP 16), Quadrupolar (e.g., SMP 27), and Pointsymmetric (e.g. SMP 10) PNs. The ratio of 
symmetric-to-asymmetric PNs\footnote{Symmetric PNs are Round and 
Elliptical PNs,
Asymmetric PNs are Bipolar, Bipolar Core, and Quadrupolar PNs. 
We do not include Pointsymmetric PNs here, for homogeneity of the
Galactic and Extra-Galactic samples.} is higher in the Galaxy than 
in the LMC. This is an indication that morphology traces the metallicity of the 
PN progenitors.

\begin{figure}
\begin{center}
\includegraphics[width=14truecm]{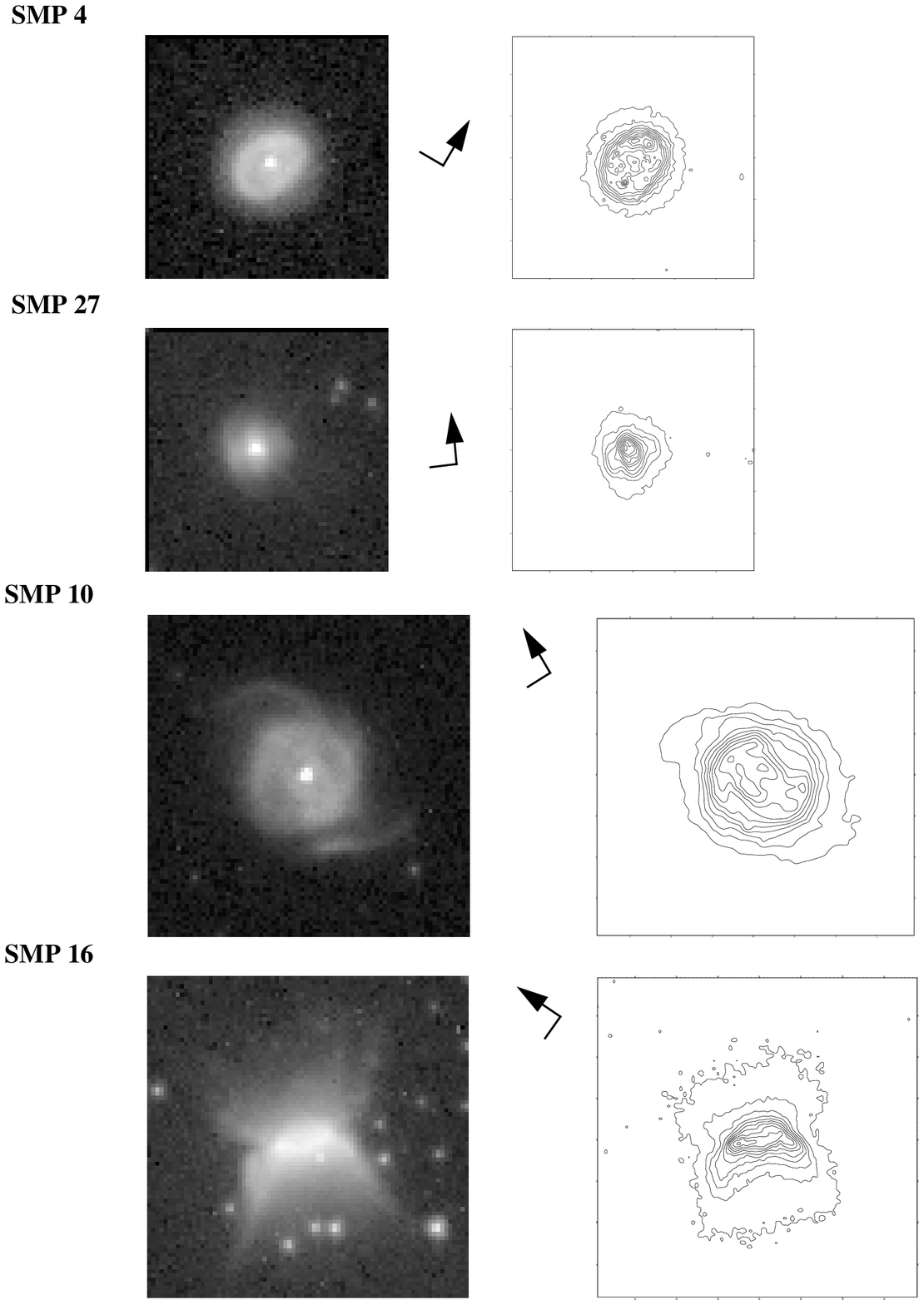}
\end{center}
\caption{Morphological sampler of LMC PNs.}
\end{figure}

In Figure 2 we show the [O III] $\lambda$5007\AA~ surface brightness evolution,
as a function of the physical radius. The LMC PNs are indicated with
different symbols, depending on their shell morphology (see Figure legend). We can see a
clear morphological separation accordingly to the evolutionary rates of the
different types. Round PNs show a slow surface brightness decline, while
Bipolar (and, in general, asymmetric) PNs evolve fast. This is true if the 
physical radius is a good measure of the dynamical time of the PNs. 
In effect, ther nebular evolution also depends moderately on the velocity of
the shell expansion (see also Shaw et al. 2001). 

By studying the chemical content of symmetric and asymmetric PNs in 
the LMC, Stanghellini et al. (2000) found that asymmetric PNs derive from 
the evolution of the youngest of the PN-producing stellar population. 
In Figure 3 we illustrate this point by showing the segregation of
the LMC PN morphological types on the basis of their neon and
sulphur abundances. The evolution of stars in the PN progenitor
mass range do not alter the abundances of sulphur and neon, thus
the plot clearly shows that there is a separation in the populations
of the progenitors of the PNs with different morphological types. 
This finding bears on the question of formation mechanisms 
for asymmetric PNs: the genesis of PNs structure should relate 
strongly to the population type, and by inference the mass of the 
progenitor star, and less strongly on whether the central star is a 
member of a close binary system, as previously believed.

\begin{figure}
\begin{center}
\includegraphics[width=11truecm,angle=0]{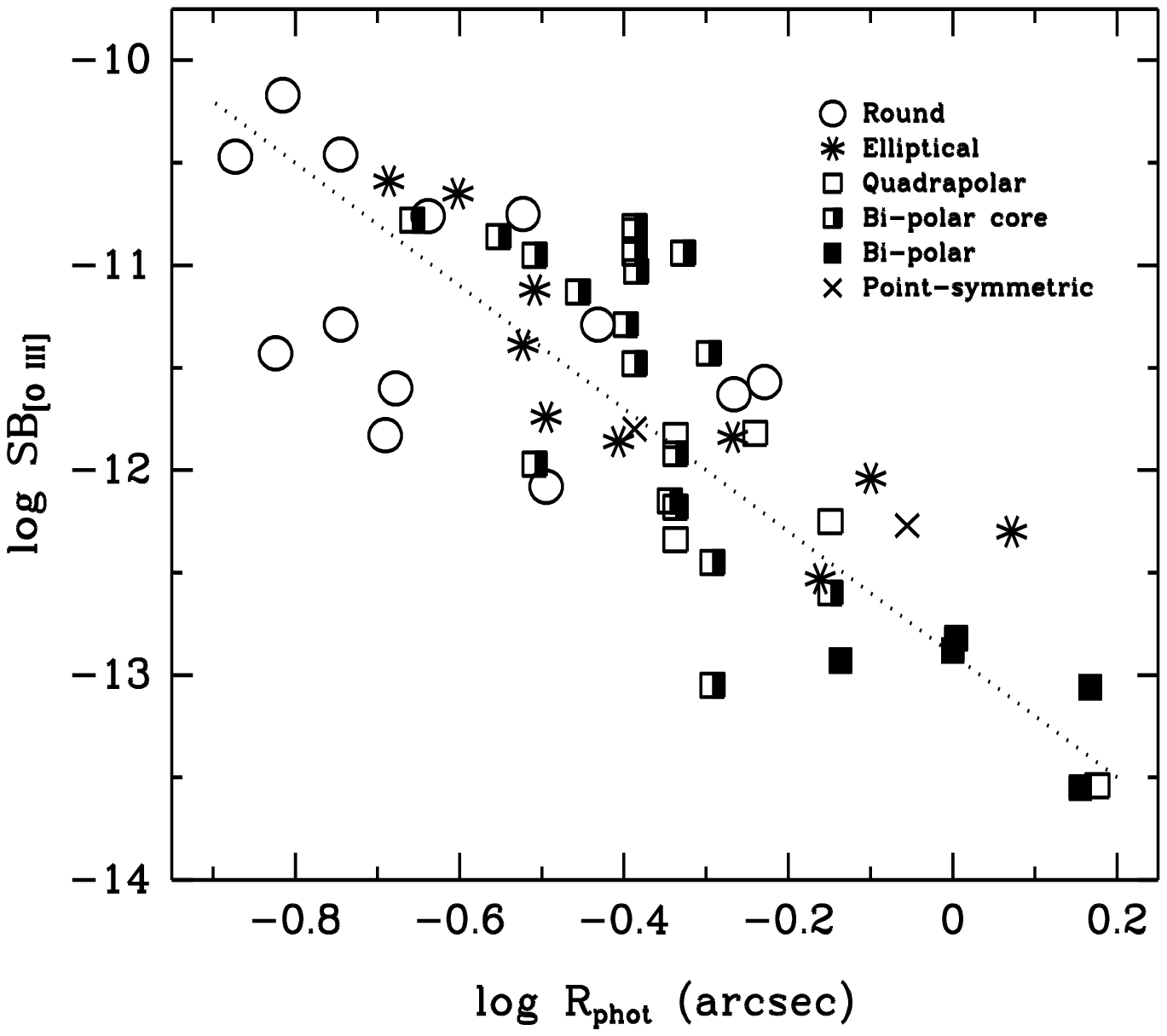}
\end{center}
\caption{The evolution of the [O III] $\lambda$5007\AA~ 
surface brightness in LMC PNs.}
\end{figure}

\begin{figure}
\begin{center}
\includegraphics[width=9truecm,angle=-90]{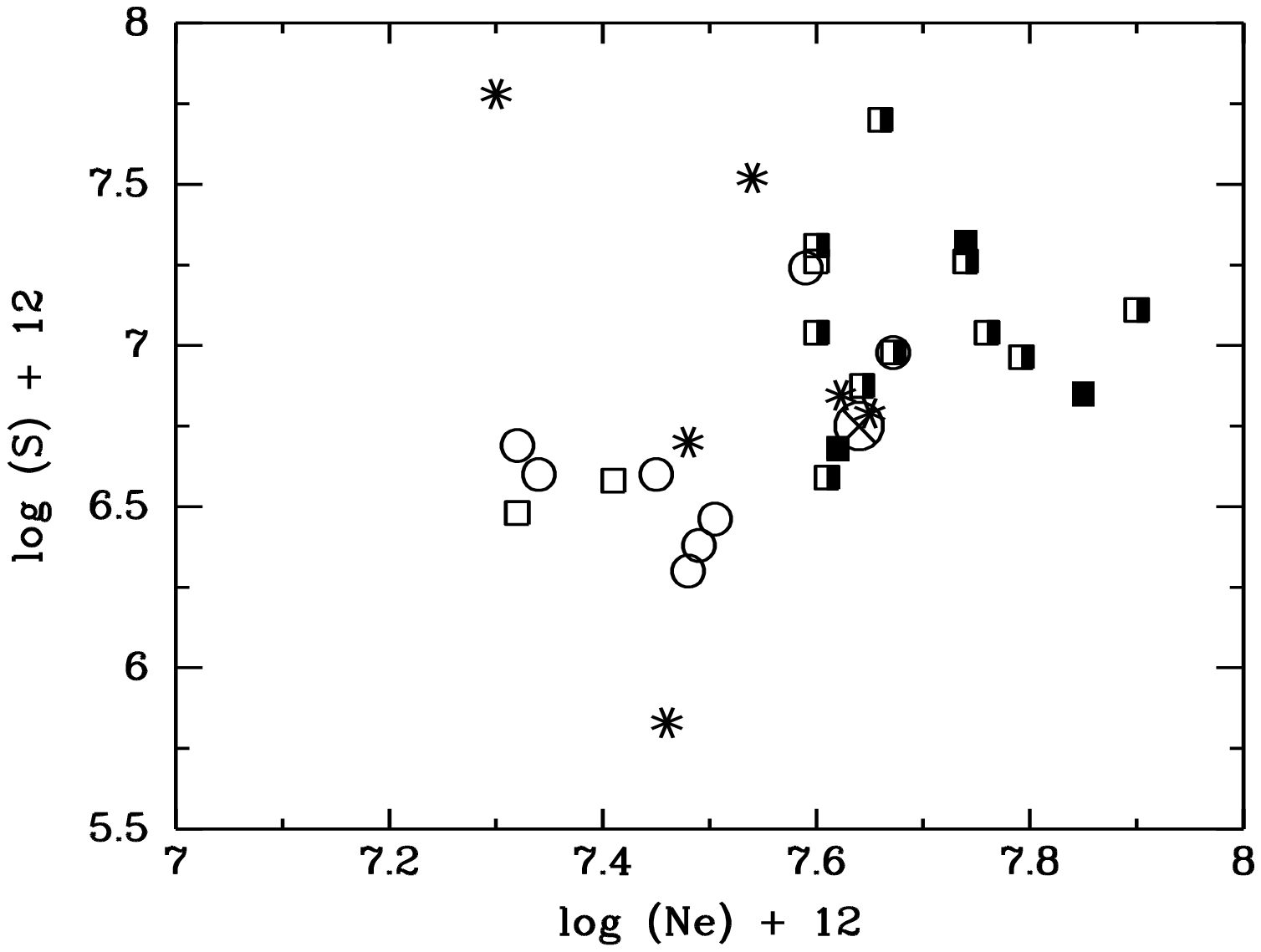}
\end{center}
\caption{Neon and sulphur abundances in different LMC PNs. Morphological
types: see legend of Fig. 2.}
\end{figure}

\section{Summary, and Future Projects}

Despite the large number of excellent studies of Galactic PNs, 
several questions on PN morphology and its formation still remain unanswered.
The study of the Magellanmic Clouds PNs has the great advantage to
produce absolute physical parameters, to constraint and probe the
existing evolutionary and hydrodynamical models. Our 
program on LMC PNs has already shown that, indeed, there is 
population and mass segregation among planetary nebulae of different
morphology, and that symmetric PNs seem to evolve slower than
asymmetric PNs. Furthermore, there are hints that the overall 
morphological distribution in a galaxian PN population depends
strongly on the galaxian type and metallicity. A follow up on these
studies bears not only on the understanding of PN morphology
and evolution, but also on the stellar populations in galaxies, and on the
planetary nebula luminosity function as a secondary indicator
of the Extra-galactic distance scale. We will strive to obtain quantitative
results on solid statistical grounds to enlighten these astrophysical
aspects.

A similar study in the Small Magellanic Cloud (SMC) is planned, to 
extend the {\it metallicity baseline} of the above findings. 
The LMC and SMC PN images acquired by Stanghellini and 
collaborators will form a database of Extra-galactic PN images that 
will far exceed in number the Galactic PNs observed with  {\it HST}, 
providing an homogeneous sample for testing the implications of 
metallicity variations in stellar evolution.

\acknowledgments

Thanks are due to J. Franco and the other organizers for the opportunity
to participate to an important celebration for the Peimberts. 
It was a pleasure to be at the Conference, and to be able to
discuss scientific and non-scientific issues with J. Franco, 
G. Garc\'{\i}a-Segura, M. Dopita, S. Oey, M. Richer, and many others.


\begin{thebibliography}

\bibitem{} Aller, L. H., \& Keyes, C. D. 1987, ApJS, 65, 405

\bibitem{}Blades, J.~C., et al.~1992, 
        ApJ, 398, L44

\bibitem{} Barlow, M.\ J.\ 1991, IAU 
Symp.\ 148: The Magellanic Clouds, 148, 291 

\bibitem{} Boroson, T. A., \& Liebert, J.~1989, ApJ, 339. 844

\bibitem{} Dopita, M. A., Ford, H. C., \& Webster, B. L. 1985, ApJ,
297, 593

\bibitem{}Dopita, M.~A., et al.~1996, 
        ApJ, 460, 320

\bibitem{} Greig, W.~E. 1972, \aap, 18, 70 

\bibitem{} Kaler, J.\ B.\ \& 
Jacoby, G.\ H.\ 1990, \apj, 362, 491 

\bibitem{} Linsday, E. M. 1955, MNRAS, 115, 248
  
\bibitem{} Manchado, A., Guerrero, M., Stanghellini, L., \&
Serra-Ricart, M. 1996, 
The IAC Morphological Catalog of Northern Galactic Planetary
Nebulae (Tenerife: Instituto de Astrof\'{\i}sica de Canarias)

\bibitem{} Manchado, A., Villaver, E., 
Stanghellini, L., \& Guerrero, M.\ ;.\ 2000, ASP Conf.\ Ser.\ 199: 
Asymmetrical Planetary Nebulae II: From Origins to Microstructures, 17 


\bibitem{} Peimbert, M.~1978, IAU Symp. 76, 
        ed. Y. Terzian, (Dordrecht: Reidel), 215

\bibitem{} Peimbert, M.~1987, RMxAA, 14, 166


\bibitem{}Peimbert, M., \& 
        Torres-Peimbert, S.~1982, IAU Symp. 83, 
        ed. D.~R. Flower, (Dordrecht: Reidel), 233

\bibitem{} Shaw, R. A., Stanghellini, L., Mutchler, M., Balick, B. \& Blades, J. C. 2001, ApJ, 548, in press


\bibitem{} Stanghellini, L., Corradi, R. L. M., \& Schwarz, H. E. 1993,
            A\&A, 276, 463

\bibitem{}Stanghellini, L., 
        Blades, J.~C., Osmer, S.~J., Barlow, M.~J., \& Liu, X.-W.~1999, 
        \apj, 510, 687

\bibitem{}Stanghellini, L., 
        Shaw, R.~A., Balick, B., \& Blades, J.~C.~2000, \apjl, 534, L167

\bibitem{} Torres-Peimbert, S.\ 
1993, RMxAA, 26, 112 

\bibitem{}Vassiliadis, E., 
        Dopita, M.~A., Meatheringham, S.~J., Bohlin, R.~C., Ford, H.~C., 
        Harrington, J.~P., Wood, P.~R., Stecher, T.~P., \& Maran, S.-P.~1998, 
        \apj, 503, 253

\bibitem{} Webster, B. L. 1969, MNRAS, 143, 113

\end{thebibliography}
\end{document}